\documentclass[pdflatex,sn-nature,twocolumn,iicol]{sn-jnl}% Style for submissions to Nature Portfolio journals

\usepackage{graphicx}%
\usepackage{multirow}%
\usepackage{amsmath,amssymb,amsfonts}%
\usepackage{booktabs}%
\usepackage{titlesec}%
\usepackage{comment}
\usepackage{xcolor}
\usepackage{hyperref}
\usepackage{nameref}

\titleformat{\section}{\normalfont\normalsize\bfseries}{\thesection}{1em}{}
\titleformat{\subsection}{\normalfont\small\bfseries}{\thesubsection}{1em}{}
\begin{document}

%\title[Article Title]{Multi XC-Functional Delta-Learning for Systematic Correction of DFT Band Gaps}
%\title[Article Title]{A materials-informed neural network for systematic correction of electronic band gaps}

\title[Article Title]{A data-driven approach for systematic correction of electronic bandgaps}

\author*[1]{Pamela C. Carvalho}
\author[1]{Asa Marif}
\author[1,2]{Erna K. Delczeg-Czirjak}
\author[3]{Sergei Simak}
\author[3]{Utkarsh Singh}
\author[1,2]{Olle Eriksson}

\affil[1]{\orgdiv{Department of Physics and Astronomy}, \orgname{Uppsala University}, \orgaddress{\street{Box 524}, \city{Uppsala}, \postcode{SE-75120}, \country{Sweden}}}

\affil[2]{\orgdiv{WISE - Wallenberg Initiative Materials Science for Sustainability}, \orgname{Department of Physics and Astronomy, Uppsala University}, \orgaddress{\street{Box 524}, \city{Uppsala}, \postcode{SE-75120}, \country{Sweden}}}

\affil[3]{\orgdiv{ Department of Physics, Chemistry and Biology (IFM), \orgname{Link\"oping University}, \city{Link\"oping}, \postcode{SE-58183}, \country{Sweden}}}

%%==================================%%
%% npj Computational Materials uses an unstructured abstract, %%
%% no subheadings, no citations, no equations.                %%
%%==================================%%

%%%%%% WORD COUNT - OURS (2026-10-27) %%%%%%%%%
%                                   | Nature Comms             | npj Comput. Mater.
% abstract:            224          | <= 200                   | <= 150
% Introduction:      1,388          |                          |
% Results:           1,842          |                          |
% Discussion:          704          |                          |
% main text:         3,934          | ~5,000 (Intro+Res+Disc)  | no strict limit
% Methods:           2,500          | typically < 3,000        | no limit given
% figure legends:  208 / 182 / 282 / 251 (Figs. 1-4)  | <= 350 each | <= 350 each
% display items:       4 figures    | up to 10                 | -
% references:         47 (paper)    | ~70 (guide)              | 60 (not strictly enforced)
% title:              11 words      | <= 15                    | <= 15, no punctuation/idioms/puns

\abstract{

We demonstrate that a single machine-learning correction, applied to bandgaps from density functional theory calculations with 
%the PBE and HSE06 
two of the most frequently used energy functionals, results in bandgaps that are close to the experimental values. This solves a classical problem of computationally efficient electronic structure methods, to give reliable bandgaps that can be compared to experiments. The employed approach relies on materials specific feature engineering of machine learning models, and is shown to have close accuracy to that of much more expensive calculations based on the GW-approximation or hybrid functionals. The findings reported here open the possibility to use efficient calculations in materials design workflows that need reliable bandgaps, with implications for investigations targeting large computational databases or for investigations that are entirely predictive in nature.
}


\maketitle

\section*{Significance statement}

%% WORD COUNT 119 - limit of 120
Semiconductors are key materials in technologies ranging from photovoltaics to power electronics, where their bandgap determines their application. Theoretical calculations can support experimental data, but the approximations normally used strongly affect predicted bandgaps. Cheap methods often underestimate bandgaps, while more accurate ones are costly. Here, we develop a data-driven workflow in which a machine-learning model learns the systematic bias between theory and experiment for two approximations. After correction, cheap and expensive calculations yield similar bandgaps that are close to experimental values. On a highly accurate benchmark, corrected bandgaps from the cheap approximation have an error only about 1.5 times that of a 1000$\times$-costlier calculation. This allows efficient materials design, where corrected cheap calculations can be a surrogate for expensive ones.

\section*{Introduction}\label{sec:intro}

Theoretical calculations have emerged as an outstanding complement to experimental investigations of materials, and are often applied to scientific questions in chemistry, physics and technology \cite{Sholl2009}.
%(here I suggest to add a reference to: D. S. Sholl and J. A. Steckel, Density Functional Theory: A Practical Introduction, Wiley 2009).
Examples can be found in investigations of superconductivity, magnetism and magnetic materials, electronics, optoelectronics and catalysis, to give a few examples, with tens of thousands of calculations \footnote{Scopus (\url{https://www.scopus.com}) lists 35,460 publications from 2025 with ``density functional theory'' in the title, abstract or keywords (searched on 27 October 2026).}
%(I got this number from SCOPUS so lets add a foot note saying that SCOPUS reports 36.000 publications 2025) 
that each year report on theoretical, materials specific, investigations. In this regard, materials for use in applications that enable sustainable energy technologies stand out, where, e.g., photo-catalysis, light-emitting diodes (LEDs), photovoltaics, and novel semiconductors rely on materials with a suitable and tuneable bandgap.

The most important theoretical methods in these investigations are based on density functional theory (DFT) \cite{Hohenberg1964, Kohn1965}, since it allows for computational efficiency to be combined with accuracy and materials specificity \cite{Ullrich2026}. However, for many of the applications mentioned above an accurate bandgap of a semiconductor or insulator is needed, something which is out of reach for standard implementations of DFT, particularly the computationally most efficient approximations to the exchange and correlation (XC) functionals \cite{Burke2012,Maurer2019, Butera2024}, like the local density approximation (LDA) and the generalized gradient approximation (GGA). Here, it is important to differentiate the Kohn--Sham (KS) bandgaps, as computed by DFT, from the fundamental (or quasi-particle) ones, defined as the difference between the ionization energy and the electron affinity. The fundamental gap generally differs from the KS bandgap by the XC derivative discontinuity, even for the exact functional \cite{PerdewLevy1983, ShamSchluter1983}. In this context, semi-local approximations such as LDA and GGA typically underestimate fundamental gaps, due to the lack of this discontinuity and to self-interaction (delocalization) errors \cite{Ullrich2026, Burke2012, Maurer2019, Butera2024, mohn_2026, zhuo_2018, setyawan_2011}. Furthermore, comparisons with experimental gaps must account for physical effects such as electron--phonon interactions, thermal expansion, pressure, and defects, which are not generally included in conventional static-lattice calculations \cite{kiselyova_2016, dong_2022}.

%The fundamental gap is the difference between the ionization energy and electron affinity, an excited-state property which is not directly accessible by the ground-state theory from the Kohn-Sham framework \cite{PerdewLevy1983, ShamSchluter1983}. In this context, DFT calculations give bandgaps that typically are too low \cite{Ullrich2026, Burke2012, Maurer2019, Butera2024, Becke1993a, Becke1993b, Stephens1994, Janesko2021, Hedin1965} and that is correlated to the well-known discontinuity in the XC potentials together with the self-interaction error \cite{mohn_2026,zhuo_2018, setyawan_2011}. Furthe, part of the systematic bias also comes from the fact that DFT calculations describe perfect crystals at zero temperature, whereas the measured gaps are dependent on temperature, pressure, defects, among others \cite{kiselyova_2016, dong_2022}. }
%Furthermore, DFT reports the Kohn--Sham gap while experimental outputs may also refer to optical gaps, and since these two quantities are not always distinguished, this can influence the comparison between experimental and DFT results.}

\begin{figure*}[h]
\centering
\includegraphics[width=\textwidth]{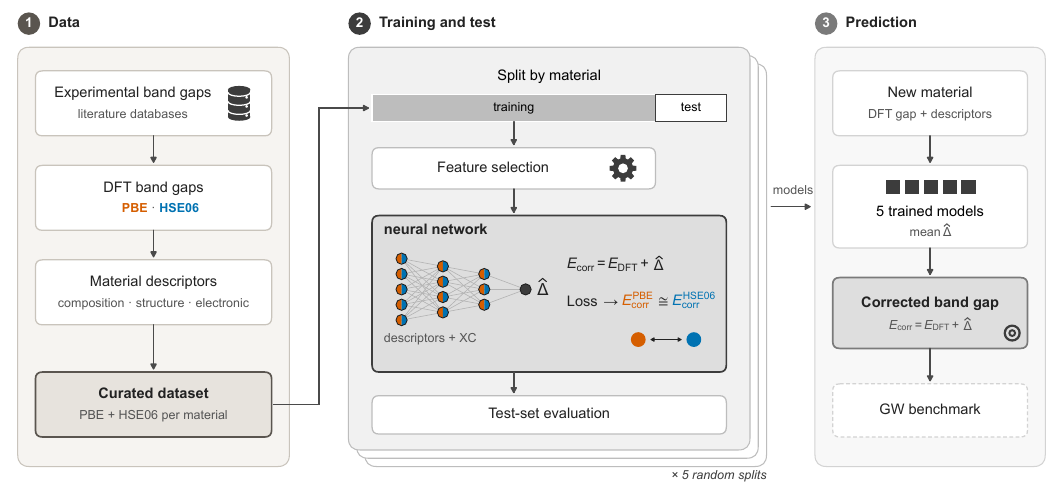}
\caption{Schematic illustration of the workflow used here. In the first stage, the pipeline starts with experimental and theoretical bandgap databases, where the DFT records include both PBE and HSE06 functionals. Descriptors spanning composition, structural and electronic features are generated, producing a curated dataset with the bandgap error between DFT and experimental results ($\Delta$), with one PBE and one HSE06 bandgap per material. In the second stage, the data is randomly split by material into train and test sets, holding out 20\,\% of the materials for testing. The train split is used for feature selection, which keeps the smallest set of descriptors that predicts $\Delta$ as well as the full set. The selected features are then used to train the neural network, whose loss encourages the corrected DFT bandgaps of the same material to agree between the PBE and HSE06 functionals. The trained model is then evaluated on the test set. This procedure is repeated for five random train/test splits. In the final stage, the five trained models are used to predict $\hat{\Delta}$ (obtained as a mean of the five predictions) and correct the DFT bandgap of new materials, including a benchmark set of materials with QP-G$_0$W$_0$ bandgaps \cite{Grossmann2026}.}
\label{fig:workflow}
\end{figure*}

For this reason, methods beyond the use of GGA and LDA have been proposed, the most noticeable being hybrid functionals 
\cite{Becke1993a, Becke1993b, Stephens1994, Janesko2021}, such as HSE06 \cite{Heyd2003, Heyd2004, Heyd2006}, and the GW approximation \cite{Hedin1965}. The former is based on a mixing of exact (Fock) exchange to the LDA or GGA functional, in particular to the Kohn-Sham potential that emerges from DFT. The GW approach is based on many-body perturbation theory, in which many-body effects and excitation energies are obtained from a self-energy that is evaluated from a screened Coulomb interaction times the Green's function of the Kohn-Sham equation. In particular, since quasi-particle energies are the energies required to add or remove one electron, this approach is able to compute the fundamental gap directly. This is a notoriously computationally expensive approach, that scales as N$^{4}$, with N being number of atoms per unit cell. There are in addition several ways to approximate this self-energy in the GW approximation (often referred to as $G_0W_0$, $GW$, and G$\hat{W}$, with the latter being the most accurate approximation \cite{Grossmann2026}, since vertex corrections to the screened Coulomb interaction are included). 

%The hybrid functionals and the GW-based calculations present, more or less, a high accuracy for band gaps} \cite{Grossmann2026}. Experience also shows that they are significantly more demanding computationally, compared to GGA and LDA based calculations.
In order to have a theoretical tool that combines the computational efficiency of conventional DFT calculations with the accuracy of GW-based methods and hybrid functionals, one has to try new approaches. In this work, we explore the use of DFT based calculations, on GGA and HSE06 functionals, combined with supervised machine-learning (ML) algorithms that are trained to estimate and predict how large error DFT-based calculations give across different XC functionals. We demonstrate that this approach reaches close accuracy to that of expensive GW-like implementations, at the fraction of the computational cost. 

%\section*{Methods-short version}\label{sec1b}
%\newpage

Although previous works in the literature have focused on predicting experimental bandgaps through machine-learning techniques \cite{gu_2006, zhuo_2018, jung_2024, debreuck2022accurate, chen2021learning}, the approach used here relies on the so called error corrected density functional theory \cite{ramakrishnan_2015, Simak2025} using a multi-XC formulation. The basic idea behind this approach is that a neural network can be trained to estimate the error a specific implementation (e.g. generalized gradient approximation - GGA or local spin-density approximation - LSDA) of density functional theory (DFT) that is unavoidable for any specific calculated property of real materials. The approach has been tested for heat of formation of compounds and alloys as well as for materials density, and crystal structures \cite{Simak2025, Singh_inprep}. In particular, we expand this framework to span the correction over multiple XC functionals, resulting in similar corrected properties across them. This method can be used for a range of physical and chemical properties of the ground state, like magnetic properties and elasticity, as well as excited state properties, like the bandgap of a solid. In this work we focus on the bandgap, and a full account of the methods is given in the \hyperref[methods]{Methods} section where we outline data-curation as well as machine learning models, feature engineering and training of the neural network.

As mentioned, GGA or LSDA implementations of DFT typically underestimate the bandgap, rendering the use of computational methods to analyze this property rather meaningless. This also prevents predictive studies based on theory, that potentially could quickly scan through many materials combinations to try to identify new functional materials. The error in GGA/LSDA based calculations is here treated as a learnable quantity: for a material with feature vector
$\mathbf{x}$ (described below) evaluated with a specific functional, $\mathrm{XC}$, one can identify a relationship between experimental ($E_{\mathrm{exp}}$) and DFT calculated bandgap ($E_{\mathrm{DFT}}$):

\begin{equation}
    E_{\mathrm{exp}} = E_{\mathrm{DFT}}^{\,\mathrm{XC}} + \Delta(\mathbf{x}, \mathrm{XC}),
\end{equation}

\noindent where $\Delta$ is the DFT error, or more precisely, the difference between the calculated gap and the experimental reference, which accounts for errors of different sources. Note that depending on level of approximation the parameter $\mathrm{XC}$ in this equation can mean GGA, LSDA, LSDA+U, a hybrid functional (as HSE06) or any of the levels of the GW approximation. In the present work we focus on GGA (in the PBE and PBE+U parametrizations) and HSE06 for $\mathrm{XC}$. The basic idea of this study is to collect $\Delta(\mathbf{x}, \mathrm{XC})$ for a large set of materials, that are characterized by a feature vector $\mathbf{x}$ (e.g. including chemical composition and crystal structure), and with supervised learning, train a neural network to make accurate estimates of the error a $E_{\mathrm{DFT}}^{\,\mathrm{XC}}$ calculation might give. We refer to this predicted error as $\hat{\Delta}$. For any DFT calculation of a new material (with unique value of $\mathbf{x}$) that is outside the training set, the ambition is then to have the ability to calculate an accurate bandgap as $E_{\mathrm{DFT}}^{\,\mathrm{XC}}
+ \hat{\Delta}$, a bandgap we below refer to as $E_{\mathrm{corr}}$, i.e., the corrected bandgap.

%To train a neural network with respect to $\Delta$, we assembled a matched experimental--theoretical dataset, pairing measured literature band gaps mined by Dong and Cole \cite{dong_2022} with calculated gaps from the Materials Project \cite{Jain2013, Horton2025}, SNUMAT \cite{kim_2020} and other sources (described in the extended methods section below), yielding a high-quality selection of 1,545 entries and 758 materials, that spans PBE, HSE06 and PBE+U functionals. For this group of materials we selected 90 \% for the training step while the remaining 10 \% were used as a test for how close $\hat{\Delta}$ could be to $\Delta$. The training set of materials were used in the training of the neural network, while the test set was left out of this training. 

To train a neural network (NN) with respect to $\Delta$, we assembled a matched experimental--theoretical dataset, pairing measured bandgaps from the curated set of Zhuo~\emph{et~al.}~\cite{zhuo_2018} and the literature data mined by Dong and Cole~\cite{dong_2022} with calculated gaps from the Materials Project~\cite{Jain2013, Horton2025}, SNUMAT~\cite{kim_2020} and other sources (described in Methods and Section S1 of the Supplementary Information, SI). This yields a high-quality selection of 1,504 entries for 752 materials, each with one PBE and one HSE06 bandgap. To assess how accurately the model predicts $\hat{\Delta}$ for entirely new materials, we randomly split the materials five times, independently, holding out 20\,\% of them (151 materials) as a test set that was never used for training, feature selection or hyperparameter choices. The final metrics reported here are the average over the five splits, given as mean $\pm$ standard deviation (sd) over the splits unless stated otherwise.
%Instead the training set was used to test how accurate the physics-informed neural network (\bc{NN}) could predict $\hat{\Delta}$ of entirely new materials and to assess how accurate band gaps one could obtain from DFT based computation combined with a \bc{NN}.
The NN used here was trained to give values of $\hat{\Delta}$ using 12--40 features, selected separately on the training set of each split (from composition,
structure and electronic descriptors, as described in the Methods section). Since the corrections for both XC functionals are trained against the same experimental reference, the corrected gaps ($E_{\mathrm{corr}}$) obtained from different functionals are brought close to each other. Besides, we also introduce an extra regularization term that slightly improves the agreement between the corrected PBE and HSE06 gaps (see Supplementary Section S8). The complete workflow of our method is
shown in Fig.~\ref{fig:workflow} and full details can be found in the \hyperref[methods]{Methods} section.

The final models use between 12 and 40 descriptors per split (described by $\mathbf{x}$ above), selected on the training set of each split by the
BGML/GBFS pipeline of Jung~\emph{et~al.}~\cite{jung_2024}, with cross-validation folds grouped by material. On average, the selected descriptors span composition
(10), structure (9), electronic structure (1), density of states (4) and the XC functional (1), where 8 descriptors are selected in every split, including the XC label. Permutation importance \cite{breiman2001random} on the test sets identifies, in average, the XC label as the dominant input of the NN (45\,\% of the total importance), followed by composition (27\,\%) and structural (17\,\%) features.
%; the Shapley Additive Explanations (SHAP) \cite{lundberg2017unified} of the random-forest baseline rank the XC label first as well. 
Besides, the orbital character of minimum/maximum conduction band, statistics concerning the average number of $p$ valence-electron in the chemical formula, along with coordination statistics, ranked highly among the most relevant descriptors. %Surprisingly, electronic descriptors, from electronic structure and density of states, were found to be less relevant for the \bc{NN}} than composition and structural ones.
The most frequently selected descriptors, family-level contributions and permutation importances 
%and SHAP summaries 
are given in the SI (Supplementary Sections S2 and S3).
As a final comment to this section, we point out that, here, PBE calculations also include PBE+U, with the addition of the Hubbard U parameter \cite{dudarev1998}. Then, every material contributes with exactly one PBE-class and one HSE06 bandgap and the XC descriptor distinguishes only PBE and HSE06. Among the PBE-class data, 18 of 752 entries include the Hubbard correction (around 2\,\%).
%As a final comment to this section, we point out that although PBE+U cases are included in the training, most of the results presented here will focus on PBE and HSE06 calculations, as the number of PBE+U entries are quite limited in number (around 2$\%$ of all entries). 

\begin{figure*}[h]
\centering
\includegraphics[width=\textwidth]{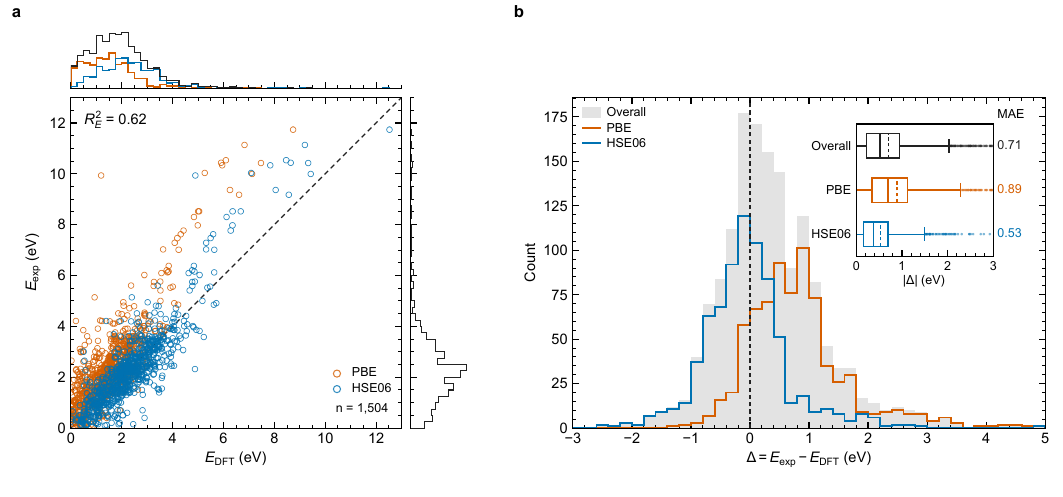}
\caption{Raw values of experimental and theoretical bandgaps for the selected dataset with \textit{n}=1,504 entries (752 materials, each with one PBE and one HSE06 bandgap). (a) Parity plot comparing the experimental ($\mathrm{E_{exp}}$) and the DFT ($\mathrm{E_{DFT}}$) calculated bandgaps, where the points are colored according to their respective XC functional (PBE, including PBE+U, and HSE06). On the edges, histograms (bins of 0.25 eV) represent the distribution of data for each axis, experiment on the right side of the plot and theory on the top. The coefficient of determination in the bandgap space ($R_{E}^{2}$) indicates the fraction of the variance of the experimental data that is explained by the theoretical one. (b) Histogram of the DFT error ($\Delta$) for the overall distribution of both sets of calculations, and for the functional (XC) specific sub-sets. The inset shows the boxplots for the overall, PBE and HSE06 $|\Delta|$ distributions (boxes: interquartile ranges; solid and dashed lines are median and mean values respectively). The values at the right side of the inset indicate the mean absolute error (MAE) for each case.}
\label{fig:baseline_error}
\end{figure*}

\section*{Results}\label{sec2}
\subsection*{Systematic bias in DFT bandgaps}\label{subsec:systematic_bias}

Comparing the experimental and theoretical bandgaps of the data collected here (training and test), there is a clear bias in the DFT predictions that correlates with the XC functional used. In Fig.~\ref{fig:baseline_error}(a), the parity plot shows a systematic underestimation for PBE calculations, while the HSE06 results lie closer to the experimental values. It can also be noted that systems with larger bandgaps present the largest errors, both from PBE and HSE06 calculations. For the total data set, that includes 1,504 entries, calculated for the two functionals, the coefficient of determination in the bandgap space (see Methods), $R_{E}^{2}$, computed between DFT ($\mathrm{E_{DFT}}$) and experimental values ($\mathrm{E_{exp}}$), is 0.62. This indicates a poor correlation between calculated and measured bandgaps, and inspection of Fig.~\ref{fig:baseline_error}(a) shows that PBE values lie significantly further away from measurements compared to HSE06 values. On the right edge of the plot we show histograms with the number of experimental bandgaps found in selected energy regions of 0.25 eV (starting from 0.1 eV as the lowest identified bandgap). The corresponding theoretical information is given at the top of the edge of the figure, for PBE and HSE06 and the total number of calculated bandgaps. 
%Both experimental set of data show that a majority of all materials have a measured band gaps at around 2 eV. 
The experimental data show that most materials have measured bandgaps between 1 and 3 eV (61\,\% with median 2.2 eV).
%indicating what is the fraction of the variance in $\mathrm{E_{exp}}$ that is explained by $\mathrm{E_{DFT}}$. Then, only around 62$\%$ of the experimental variance is captured by DFT. 

In Fig.~\ref{fig:baseline_error}(b), the histogram of $\Delta$ shows a distribution centered slightly above zero, with most values between $-2$ and $+3$ eV and a long tail towards positive errors (DFT underestimates the bandgap). The differences for the two XC-functionals are evident here, where PBE is found to be shifted significantly towards positive errors while HSE06 sits roughly around zero, with a tail extending primarily to negative errors resulting in a small negative $\Delta$ average. The mean absolute error (MAE, see inset) reflects the XC performance: HSE06 predicts bandgaps more accurately than PBE, as is already established in other works \cite{kim_2020, mohn_2026}. Interestingly, both functionals used have a bias, HSE06
with a negative value of the average $\Delta$-value and PBE with a clear positive value of the average $\Delta$-value. However, the distribution of $\Delta$ is for both functionals such that a single shift of a calculated bandgap would not provide an accurate correction to the DFT error, since this error varies significantly from material to material. This reinforces the need for ML models that can learn and correct it.

We end this section with a note that some materials have more than one literature measurement listed of the bandgap (for 249 of 752 materials). For these systems, the median sd of the reported gaps is 0.10 eV and the median spread between the lowest and highest values is 0.20 eV, indicating that part of the residual error shown in Fig.~\ref{fig:baseline_error}(b) is intrinsic to the experimental reference data (see Supplementary Section S1).

\subsection*{PBE and HSE06 reach similar accuracy after correction}\label{subsec2}

\begin{figure*}[tb]
\centering
\includegraphics[width=\textwidth]{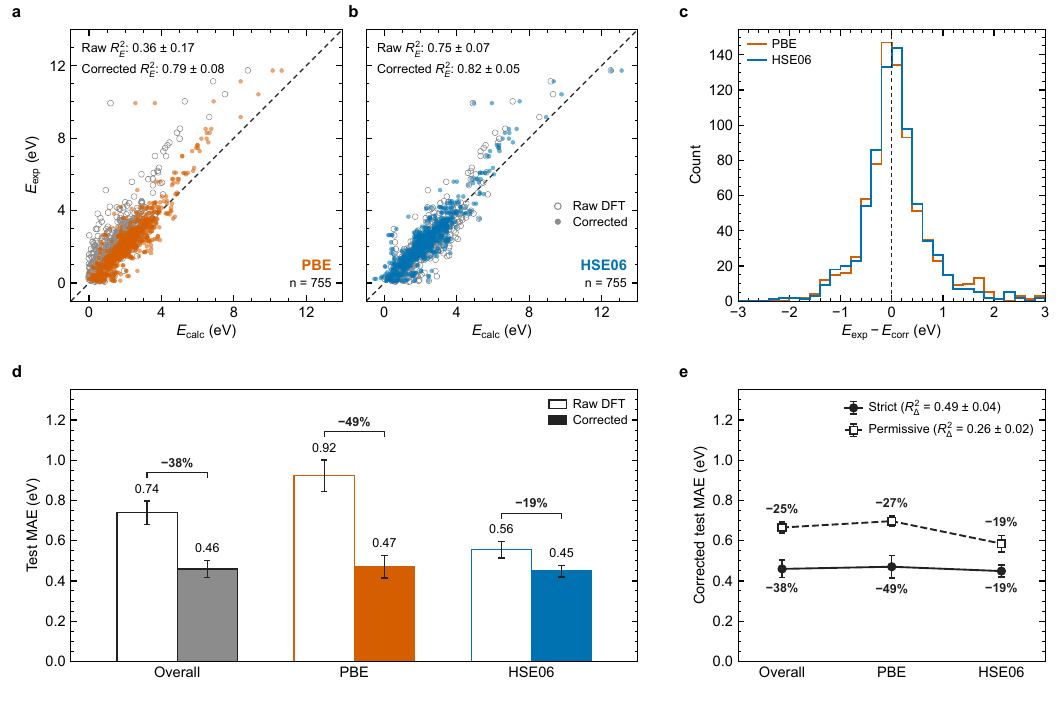}
\caption{Bandgap correction and metrics in the test sets of the five splits pooled together (\textit{n} = 1,510 entries, 496 distinct materials and 755 per functional in (a) and (b)). Parity plot of the experimental against the calculated bandgap (E$_\mathrm{calc}$), which is either the corrected (E$_{\textrm{corr}}$) or raw (E$_{\textrm{DFT}}$) DFT bandgap values for PBE (a) and HSE06 (b). Raw values, i.e., DFT bandgaps before correction, are represented as grey open circles and corrected points are colored according to their respective exchange-correlation (XC) functional. The coefficients of determination in the bandgap space ($R_{E}^{2}$) are given for the raw DFT values and the corrected values, both computed against the experimental bandgaps (mean $\pm$ standard deviation over the five splits).
(c) Histogram of the error between the experimental and corrected bandgap (E$_{\textrm{exp}}$ - E$_{\textrm{corr}}$) for both PBE and HSE06 functionals (bins of 0.2 eV); (d) Test MAE for raw and corrected values (mean over the five splits), overall and per XC. The reduction percentage after correction, relative to the raw DFT values, is indicated above the brackets for each case; (e) Corrected test MAE (mean over the five splits) of the strict and permissive cohorts (lines are a guide to the eye). The reduction percentage of each point relative to the raw DFT values is indicated for each case. The coefficient of determination in $\Delta$ space ($R_{\Delta}^{2}$, mean $\pm$ standard deviation over the five splits) is computed with the true and predicted DFT errors and gives the fraction of the variance of $\Delta$ explained by the model. In (d) and (e), error bars are the standard deviation over the splits.}\label{fig:correction}
\end{figure*}

In Fig.~\ref{fig:correction}(a-b), the corrected bandgaps (E$_{\mathrm{corr}}$) and the DFT bandgaps ($E_{\mathrm{DFT}}$) are compared to experimental bandgaps ($E_{\mathrm{exp}}$). Note that this is done for the test sets of materials, that were not used in the training of the model, pooled over the five splits. The figure shows data for the two considered XC-functionals separately, with PBE in (a) and HSE06 in (b).
The correction increases the overall $R_{E}^{2}$ score, in the test set, from $0.55 \pm 0.12$ for the raw data to $0.81 \pm 0.06$ after correction (considering both XC functionals), as the corrected values (filled circles) lie considerably closer to the diagonal of the plot and therefore with a significantly better agreement with experimental values. It can also be seen from the plots that corrected data from the PBE and HSE06 calculations group equally close to the diagonal of the parity plots, and perform more or less equally well.

The distribution of the corrected error ($\mathrm{E_{exp} - E_{corr}}$) in Fig.~\ref{fig:correction}(c) shows an interesting behavior. Unlike the raw uncorrected values of $\Delta$, the distributions for corrected PBE and HSE06 data now behave similarly. Both sets of data are also centered at a value where experiment and theory are in almost perfect agreement. Since the distribution of materials shown in the plot has become more narrow, compared to the data of uncorrected bandgaps shown in Fig.~\ref{fig:baseline_error}(b), that strongly suggests that the model is also accounting for corrections that are material-dependent instead of only removing the bias from the XC functional. Fig.~\ref{fig:correction} demonstrates that corrections shift the entire cohort of test materials with an average deviation from experiments that is close to zero, and in addition reduces the spread of data to be in better agreement with measurements. 

This trend is also reflected in the mean absolute error (MAE) shown for the test sets of materials in Fig.~\ref{fig:correction}(d). Remarkably, although the percentage reductions in MAE differ, the final values are similar: 0.47 $\pm$ 0.06 eV and 0.45 $\pm$ 0.03 eV for PBE and HSE06, respectively. While PBE shows a large MAE reduction of 49$\%$, the reduction for HSE06 is only 19$\%$. Therefore, a corrected PBE calculation, with the clear advantage of superior computational efficiency, reaches the same accuracy as a corrected hybrid functional value and a better accuracy than uncorrected HSE06 data. The information shown in Figs.~\ref{fig:correction}(a-d) are the central results of this study.

It is important to note that other works have also used data from different XC functionals to improve the prediction of experimental bandgaps. De Breuck \emph{et al.}~\cite{debreuck2022accurate} showed that learning the correction to the DFT bandgap is the best of several approaches, while Chen \emph{et al.}~\cite{chen2021learning} reported a 22--45\,\% reduction of the MAE of predicted experimental bandgaps.
Notably, in both cases, the reduction is measured against a model trained only on experimental data, and not against the raw DFT bandgaps, as is done here. In this work, we focus on correcting both PBE and HSE06 bandgaps with a single model, trained against the same experimental reference, so that the corrected bandgaps of the two functionals agree with each other and with experiment.

To test the model's sensitivity, a larger subset of data was also considered, including more experimental records with higher noise and materials calculated with a single functional (see Supplementary Section S1). This cohort is denoted \textit{permissive}, while the main, smaller subset used throughout this study, that is shown in Fig.~\ref{fig:baseline_error} and Fig.~\ref{fig:correction} (a-d), is denoted as \textit{strict}. Accordingly, the \textit{permissive} cohort was also digested by the feature selection workflow and another model was trained on the chosen descriptors, using the same architecture and training settings as the NN (see Methods). 
%with the consistency weight $\lambda$ chosen for each split by the same cross-validation procedure. 
A qualitative comparison between model performance for the \textit{strict} and \textit{permissive} cohorts is shown in Fig.~\ref{fig:correction}(e). Although the \textit{permissive} case still shows an overall MAE reduction of 25$\%$, the model performs systematically worse in all five splits than in the \textit{strict} case, with a coefficient of determination in the DFT error space (see Methods) $R_{\Delta}^{2}$ of only 0.26 (against 0.49 for the strict cohort).
%It is worth noting that the $R_{\Delta}^{2}$ score stated here refers to the model, using the true and predicted values of $\Delta$ to be computed. 
Because the strict cohort's data are classified with high confidence as correctly portraying experimental bandgaps, this subset is less noisy than the \textit{permissive} case, helping the model achieve higher performance. These results therefore emphasize the impact of high-quality data over quantity.

%In addition to the \bc{NN}, a random forest (RF) \cite{breiman2001random} and a linear regression model have also been trained on the data to compare performance (see Supporting Information). Although the MAE and $R_{\Delta}^{2}$ metrics are similar for the \bc{NN} and RF models, the latter shows greater overfitting, indicating that the \bc{NN} generalizes better to unseen data at similar accuracy. To investigate how relevant the multi-XC approach is to the prediction, the \bc{NN} and RF were also trained on single-XC data for comparison with the multi-XC model. While the RF's MAE is unaffected by the presence of multiple XC functionals in the training data, the \bc{NN} shows a reduced MAE when trained on multi-XC data compared to the single-XC case. 

In addition to the NN, a random forest (RF) \cite{breiman2001random} and a linear regression model have also been trained on the data to compare performance (see Supplementary Section S4). Although the MAE and $R_{\Delta}^{2}$ metrics are similar for the NN and RF models, the latter shows a larger difference between training and test errors (0.31 eV for the RF against 0.19 eV for the NN), which reflects that its deep trees fit the training data closely. To investigate how relevant a multi-XC approach is to the prediction, the NN and RF were also trained on single-XC data for comparison with the multi-XC model. While the RF's MAE is unaffected by the presence of multiple XC functionals in the training data, the NN shows a reduced MAE for HSE06 (an average of 2\,\% across all five splits) when trained on multi-XC data compared to the single-XC case, where for PBE the multi-XC can be slightly worse. In the permissive cohort, in contrast, multi-XC training reduces the NN's MAE for both PBE and HSE06 (by 1\,\% on average), while the RF also remains unchanged (see Supplementary Section S5).

\subsection*{Corrected gaps approach GW accuracy at lower cost}\label{msubsec1}

\begin{figure*}[tb]
\centering
\includegraphics[width=\textwidth]{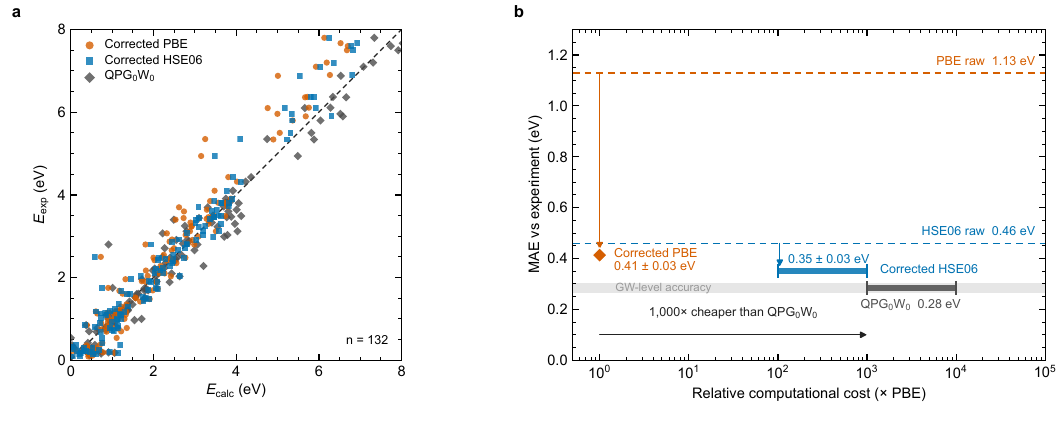}
\caption{Comparison between the corrected PBE and HSE06 bandgaps and QP-G$_0$W$_0$ calculations against experiment, with \textit{n}=132 entries (experimental bandgaps between 0 and 8 eV). The materials considered here, with PBE, HSE06 and GW calculations, were taken from Ref.~\cite{Grossmann2026}. The five split models were used to predict $\hat{\Delta}$ and correct the PBE and HSE06 bandgap values, averaging their predictions. (a) Parity plot of the calculated bandgap (E$_\mathrm{calc}$), which is either the corrected results (E$_{\textrm{corr}}$) or QP-G$_0$W$_0$, against experimental values. (b) Mean absolute error (MAE) with respect to experiment as a function of the relative computational cost (relative to that of PBE, defined as a factor of 1). For the corrected bandgaps, the MAE is the mean $\pm$ standard deviation of the MAEs of the five split models. Raw values (dashed line), QP-G$_0$W$_0$ calculations and corrected bandgaps (orange lozenge for PBE and blue bar for HSE06) are shown with their corresponding computational cost (vertical arrows mark the reduction of the MAE by the correction, and the horizontal arrow the cost ratio between corrected PBE and the lower end of the QP-G$_0$W$_0$ range). To estimate the computational cost, we followed Ref.~\cite{das_2025} and assigned orders of magnitude of $10^{2}$--$10^{3}\times$ for HSE06 and $10^{3}$--$10^{4}\times$ for QP-G$_0$W$_0$, relative to PBE. The light grey band marks the QP-G$_0$W$_0$ MAE ($\pm 0.02$ eV) as a guide to GW-level accuracy.
}\label{fig:gw_comp}
\end{figure*}

To verify the performance of the correction against highly accurate methods, we consider materials calculated with PBE, HSE06 and GW, from Ref.~\cite{Grossmann2026}. The five split models have been used to predict $\hat{\Delta}$ for the PBE and HSE06 functionals, averaging their predictions, and the resulting corrected bandgaps are compared against experiment, together with the most accurate calculation available (QP-G$_0$W$_0$). In this comparison, 132 materials are included, with experimental bandgaps between 0 and 8 eV. In Fig.~\ref{fig:gw_comp}(a), the parity plot shows the corrected gaps and GW calculations against the experimental reference. The PBE and HSE06 corrected values collapse toward the diagonal with similar behavior, and all points lie very close to the QP-G$_0$W$_0$ results. With the exception of a few outliers, all theoretical data in Fig.~\ref{fig:gw_comp}(a) track closely the experimental bandgaps.

Among these materials, 
%61 belong to the strict cohort (same composition and crystal structure), and 
36--40 of them are in the training set of each split, while the other 92--96 were not used to train that split's model. For these materials, the corrected PBE and HSE06 bandgaps of that split's model deviate from experiment by $0.48 \pm 0.04$ and $0.40 \pm 0.03$\,eV, respectively, against 0.41 and 0.35\,eV for all 132 materials in Fig.~\ref{fig:gw_comp}. Detailed metrics for materials unseen by the models can be found in the SI (see Supplementary Section S6).

In Fig.~\ref{fig:gw_comp}(b), the MAE is plotted as a function of the relative computational cost. The corrected PBE MAE value reduces from 1.13 to $0.41 \pm 0.03$ eV, for this selection of materials, while the HSE06 MAE is reduced from 0.46 to $0.35 \pm 0.03$ eV. The behavior discussed in Fig.~\ref{fig:correction}(d) is hence also found in Fig.~\ref{fig:gw_comp}, where PBE shows a larger reduction than HSE06, bringing the two functionals to comparable MAE values, with a difference of only 0.06 eV between them. The MAE for the GW calculation, according to Ref.~\cite{Grossmann2026}, is 0.28 eV (an average between QP-G$_0$W$_0$ with and without spin-orbit coupling). 

The estimation of computational cost was based on the Supplemental Material of Ref.~\cite{das_2025} computed for the 2-atom primitive cell of InSb. The HSE06 and GW calculations are represented as bars, since computational time varies according to material size. Defining the PBE calculation as the baseline and considering the lower end of the GW cost range, the corrected PBE offers a calculation that is 1000$\times$ cheaper with an error around 1.5 times that of the QP-G$_0$W$_0$. Note that this factor can be even larger, as QP-G$_0$W$_0$ can be more expensive than other GW methods \cite{Grossmann2026} and the calculation cost depends on the system size.

The lower cost of the corrected PBE calculations can be of importance for the screening of materials with potential application in semiconductor-based technologies. As an example, we consider experimental bandgaps between 1.0 and 1.5\,eV, where the Shockley--Queisser limit for single-junction solar cells is close to its maximum~\cite{Shockley1961, Ruhle2016}. Among the 73 materials with experimental bandgaps in this window and an out-of-sample prediction (using split models that were not trained on them), 
%raw PBE places only 16 inside the window,
not only the corrected PBE bandgaps place 31 of them inside the window, but the values are also closer to experiment. Similarly, we have also corrected the bandgaps of materials without an experimental reference, and highlight those whose corrected bandgaps lie within the 1.0--1.5\,eV window (Supplementary Section S7). Therefore, a corrected PBE calculation can be used to filter candidate materials for an application window before performing more expensive hybrid or GW calculations, or even for a direct comparison to experimental observations.

\section*{Discussion and Conclusion}\label{sec3}

Our results show that training a neural network to predict the DFT error of electronic bandgaps, $\hat{\Delta}$, for two XC functionals, yields corrected bandgaps with comparable MAE for the PBE and HSE06 functional (with a difference of only 0.02\,eV between them, on the test set). Applied to the QP-G$_0$W$_0$ benchmark of Gro{\ss}mann~\emph{et~al.}~\cite{Grossmann2026}, the correction brings bandgaps close to a GW-level of the MAE, at a computational cost of PBE calculations (we obtain corrected PBE bandgaps with a MAE of 0.41\,eV compared to 0.28\,eV for QP-G$_0$W$_0$ on the filtered set). This reduction in computational effort is important for several aspects of materials theory geared towards bandgap engineering, not least high-throughput screening, where hybrid functionals and GW calculations remain too time consuming to offer an efficient route. A PBE calculation plus the present correction can be a practical surrogate when hybrid/GW accuracy on the gap is desired but full hybrid or many-body calculations are not affordable. To make this possible for other users, the trained models are available through a web application (see \url{https://correctingaps.streamlit.app/} and Data availability).

Unlike for HSE06, the mean signed error of PBE is almost as large as its MAE, indicating that the PBE error is dominated by a systematic underestimation of the bandgap.
%, where the absolute mean signed error accounts for about 91\% of the PBE MAE, compared with only about 11\% for HSE06. 
After correction, the residual distributions for the two functionals become similar, consistent with the model removing much of that PBE bias while learning material-dependent corrections for HSE06. Because the HSE06 error is largely scattered around experimental references, a rigid shift cannot suffice and a material-dependent model for $\hat{\Delta}$ is required. Experimental references can also carry intrinsic scatter from sample preparation, measurement conditions, equipment uncertainties, etc. Here, where more than one experimental literature value is available (for 249 materials), the median sd is 0.10\,eV, so part of the residual after correction might be attributed to the reference data.

The multi-XC training objective, using both PBE and HSE06 for the training, allows a single model to correct both functionals, and training both against the same experimental reference brings their corrected gaps close to each other. Relative to single-XC models, the NN improves test MAE for HSE06, while for PBE it is slightly worse. In the permissive cohort it improves for both functionals, while a random forest is almost insensitive to the multi-XC setting. At a similar accuracy to that of a random forest, the NN thus gains from training on both functionals, while the random forest does not. Although switching the regularization term off leaves the accuracy almost unchanged, the disagreement between the corrected PBE and HSE06 gaps of the same material slightly increases by 0.01 eV (see Supplementary Section S8). Then, the regularization term here acts mainly on the agreement between predictions across different XC functionals instead of accuracy. The strict versus permissive comparison adds a point on the data side. The permissive model, trained on a larger but noisier set, still reduces MAE, yet underperforms the strict model by a wide margin (test $R_{\Delta}^{2}$ of 0.26 versus 0.49). High-confidence experimental labels matter more here than a larger number of samples.

%The multi-XC training objective, using both PBE and HSE06 for the training, further helps the accuracy of the \bc{NN}, as the consistency constraint encourages corrected gaps for the same material to agree across functionals. Relative to single-XC models, the \bc{NN} improves test MAE, whereas a random forest is insensitive to the multi-XC setting. Together with the smaller train--test gap of the \bc{NN} compared with the forest, this supports the physics-informed formulation beyond a generic regressor at similar absolute accuracy. The strict versus permissive comparison also reinforces the same point on the data side. The permissive model, trained on a larger but noisier set, still reduces MAE, yet underperforms the strict model by a wide margin (test $R_{\Delta}^{2}$ of 0.23 versus 0.57). High-confidence experimental labels matter more here than a larger number of samples.

It should be clearly stated that a few limitations follow for the model. One of them is the PBE+$U$ entries which are few ($\approx$ 2\%) and are treated as PBE. The model should therefore be applied to plain PBE bandgaps, and its use for PBE+$U$ calculations would require more PBE+$U$ training data. Also, the HSE06 values from SNUMAT are one-shot hybrid calculations on PBE starting points, and may differ from self-consistent hybrid calculations. Furthermore, the experimental data used here do not report whether the measured bandgaps are direct or indirect. Although we included this information as a descriptor computed from the band structure, the feature selection workflow never selected it.
Besides, wide-gap compounds produce some of the largest DFT errors, such as MgO and CaO, lying outside of the $\hat{\Delta}$ range of approximately $-1.7$ to $+2.4$ eV (the range kept in training by the outlier filter, see Methods). Therefore, the model targets smaller-gap materials, where wide-gap cases might lead to inaccurate inferences. The test sets are not filtered, so the reported test errors include such materials.

Overall, treating $\hat{\Delta}$ as a learnable, XC-dependent target of a neural network opens the possibility to use the abundance of available and efficient DFT calculations to results that have experiment-quality for the bandgaps.
The approach is therefore well suited to materials design workflows that need reliable bandgaps at the scale of large computational databases or for investigations that are entirely predictive in nature, where one may obtain bandgaps with minimal computation effort, of systems that have not yet been synthesized.

\section*{Materials and Methods}\phantomsection\label{methods}

\subsection*{Data curation}

Experimental bandgaps were taken from the ChemDataExtractor database of Dong and Cole~\cite{dong_2022} and from the curated experimental set of Zhuo~\emph{et~al.}~\cite{zhuo_2018}, the latter in the version of the Matbench benchmark~\cite{Dunn2020} with Materials Project (MP) identifiers as distributed through matminer~\cite{Ward2018Matminer} (Kingsbury curation~\cite{kingsbury_2022}). Dong and Cole comprises 100,236 records mined from thousands of articles and previously published sources, so there is partial overlap with the smaller Zhuo/Kingsbury set. In particular, the curated BandgapDB of Kiselyova~\emph{et~al.}~\cite{kiselyova_2016} was integrated into Dong and Cole and appears in our pipeline as a high confidence tier.

Because most Dong and Cole records come from a natural language processing (NLP) parser with an associated confidence score, we assign each record to a quality tier, reflecting how strongly we trust the extracted value. BandgapDB and Zhuo/Kingsbury entries form the \textit{high} tier. NLP records with confidence $\ge 0.65$ are labeled \textit{medium} if the source text is free of computational or non-bulk context flags, \textit{low} otherwise and remaining records are \textit{poor}. Extra filters on formula validity, bandgap range and multiple measurements were also applied (see Supplementary Section S1).

To link experiment to theory, MP identifiers were used when already present or obtained by formula-ranked queries to the Materials Project~\cite{Jain2013}, preferring low energy-above-hull entries.
%($E_{\mathrm{hull}}\le 0.2$\,eV/atom).
Unmatched formulas were pursued through SNUMAT~\cite{kim_2020} and OPTIMADE providers \cite{Evans2024OPTIMADE, Andersen2021OPTIMADE} (including Alexandria \cite{Cavignac2026Alexandria} and OQMD \cite{saal2013oqmd}).
%and JARVIS \cite{Choudhary2020JARVIS})
When the MP queries had no match, SNUMAT supplied both the bandgaps and the crystalline structure. Among several MP entries of the same formula, the stable entry was preferred, then an experimentally known one, then the lowest energy above hull. For each matched material we retained at most one theoretical gap per XC functional, where 
%Materials Project tasks were deduplicated by functional; 
SNUMAT provided the HSE06, and for a few cases, PBE gaps. SNUMAT entries were linked to MP entries through their Inorganic Crystal Structure Database (ICSD) \cite{zagorac2019icsd} identifiers. SNUMAT PBE gaps of compounds containing V--Zn or Ce, for which SNUMAT applies a Hubbard $U$, were labelled PBE+$U$, and the remaining PBE and PBE+$U$ gaps were taken from MP.

After querying the different databases available, our matched set contained several XC functionals (PBE, HSE06, PBE+$U$, mBJ, and others). We kept the three most abundant ones (PBE, HSE06 and PBE+$U$) for the remainder of the study, with PBE+$U$ treated as PBE. Each material contributes at most one PBE-class gap, with the plain PBE value preferred when both exist. We then defined two nested cohorts: a \textit{strict} subset (1,504 rows; 752 materials), restricted to \textit{high} experimental records for which each material has one PBE-class and one HSE06 gap, and a \textit{permissive} superset (4,874 rows; 3,619 materials), which additionally retains the remaining lower-confidence tiers and materials covered by only a single functional. In the strict cohort used for the main results, theoretical gaps come from MP and SNUMAT, where OPTIMADE matches appear only in the permissive set.

The strict cohort was split by reduced formula using stratified group
$K$-fold partitioning ($K=5$, stratified by XC functional) \cite{kohavi1995study} as implemented in scikit-learn \cite{scikit-learn}, holding out one fold as the
test set so that all XC entries for a given material appear in only one
split. This was repeated with five random seeds, giving five independent splits, so that a material can be in the test set of several splits. On the training fold only, outliers were removed through rows with $\Delta$ outside
$[Q_1 - 1.5\,\mathrm{IQR},\, Q_3 + 1.5\,\mathrm{IQR}]$, where $Q_1$ and $Q_3$ are the first and third quartiles of the training $\Delta$ values and $\mathrm{IQR} = Q_3 - Q_1$ is the interquartile range \cite{tukey1977exploratory} (approximately $-1.7$ to $+2.4$ eV), and any training material left with
fewer than two XC functionals was dropped, yielding 1,102--1,118 training and 302
test rows (551--559 / 151 materials), while the test sets were not filtered. A similar procedure has been done on the permissive cohort, without the two-functional requirement (2,734--2,754 training and 724 test materials).

\subsection*{Feature engineering}

Different families of features were generated through the featurizers available from matminer, where the features are grouped by composition, structure, density of states (DOS), and band structure (BS). For the composition, the chemical formula is used to generate several descriptors based on statistical operations across elemental properties, together with stoichiometry and orbital-based descriptors. The structural features were computed on the corresponding Crystallographic Information File (CIF) of each material, obtained from the theoretical data sources. The DOS and BS were retrieved from MP, when available, where the featurizers generated descriptors from these objects. Five descriptors are categorical: the XC functional, the crystal system, whether the bandgap is direct or indirect (from the BS), and the orbital character of the conduction-band minimum and of the valence-band maximum (from the DOS). Descriptors that essentially reproduce the theoretical bandgap were excluded by design, so that the model learns the correction from material descriptors instead of $E_{\mathrm{DFT}}$ itself. Missing descriptors were filled with the median of the training set.

Since the generated number of features was large (+700), a Gradient Boosted and Statistical Feature Selection (GBFS) workflow \cite{jung_2024} has been used on the training set of each split, to reduce overfitting and to limit the noise that uninformative descriptors might introduce into the predictions. The GBFS workflow combines statistical filters to remove features that are constant or correlated with others with a feature selection procedure based on the ranking of features using a Light Gradient Boosting Machine (LightGBM) model \cite{ke2017lightgbm}, evaluated by five-fold cross-validation with folds grouped by material. 
%Instead of the recursive cut on a single random split of the original workflow, whose result depends strongly on that split, 
Here, the number of descriptors is the smallest whose cross-validated error lies within 1.5\,\% of the lowest. Also, while the XC functional was not forced into the selection, it was selected in every split. Between 12 and 40 descriptors were selected per split, on average 10 from composition, 9 from structure, 4 from DOS, 1 from BS and the XC functional. Among those, 8 descriptors are common to all splits.

The permissive cohort was also included in the workflow yielding a different selection of 16 to 28 features per split, on average 10 from composition, 9 from structural, 3 from DOS, none from BS and the XC functional. Among those, 12 descriptors are common to all splits. The materials included in the test set were never used for the feature selection pipeline. More details about the generated features and the final selection can be found in Supplementary Section S2.

\subsection*{Machine-learning models}

The main model used here was a neural network (NN) trained on the DFT error ($\Delta$). Besides minimizing the standard mean-squared error (MSE) on $\Delta$, a regularization term was added to the loss function, which penalizes differences between the corrected bandgaps of the same material across XC functionals, where

\begin{equation*}
\begin{aligned}
\mathcal{L}
=
\mathcal{L}_{\mathrm{MSE}}(\hat{\Delta},\Delta)
+ \\
&\lambda\,
\frac{1}{M}\sum_{n \in \mathcal{M}}
\left[
\hat{E}_{\mathrm{corr},\,n}
-
\bar{E}_{\mathrm{corr},\,m(n)}
\right]^{2}
\end{aligned}
\end{equation*}

\noindent where \(m(n)\) is the material group of row~$n$, \(\bar{E}_{\mathrm{corr},\,m(n)}\) is the mean corrected gap over XC entries of material~$m$, $\mathcal{M}$ is the set of the $M$ training rows that belong to materials with at least two XC entries and 

\begin{equation*}
\hat{E}_{\mathrm{corr},\,n} = E_\mathrm{DFT}^{\,\mathrm{XC}_{n}} + \hat{\Delta}(\mathbf{x}_{n},\mathrm{XC}_{n}).
\end{equation*}

Since the PBE and HSE06 entries of a material share the same experimental reference, the MSE already favors agreement between their corrected gaps and therefore the regularization term only increases the weight given to this agreement.
The features were standardized and categorical descriptors were included through one-hot encoding. Here, $\Delta$ was standardized with the training mean and sd in $\mathcal{L}_{\mathrm{MSE}}$, while the regularization term is evaluated in eV$^2$. The network is a Multilayer Perceptron (MLP) \cite{rumelhart1986learning} with two hidden layers of 256 and 128 units, ReLU activations and the AdamW optimizer (learning rate $5\times10^{-4}$, weight decay $10^{-4}$, gradient clipping at 5). These settings were kept fixed for all splits and cohorts. The regularization weight $\lambda$ was chosen for each split by five-fold cross-validation grouped by material on the training set, from the grid $\{0, 0.1, 0.5, 1, 2, 5, 10, 20\}$ (selected values 5--20 for the strict and 0.1--1 for the permissive cohort). Including $\lambda = 0$ in the grid allows a direct comparison with a network without this term (see Supplementary Section S8). The final model of each split is the mean prediction of an ensemble of 10 networks trained with different random seeds, each early-stopped on the mean absolute error of an inner validation set of 15\,\% of the training materials. In the permissive cohort, materials with a single functional contribute only to $\mathcal{L}_{\mathrm{MSE}}$.

Besides, an ordinary least squares (OLS) model and a random forest (RF) have also been trained on the same data to compare performances. A randomized search with material-grouped cross-validation was used to select the RF hyperparameters. For the OLS case, the XC functional was one-hot encoded and the features were standardized. Remaining training details are presented in the SI (see Supplementary Section S8).

We also compared the NN and the RF by training single and multi-XC models. In the multi-XC setting, one model is trained on all functionals together (for the NN, including the regularization term). In the single-XC setting, a separate model is trained using only the rows of one functional (PBE-only or HSE06-only), with the same architecture and hyperparameters, and is evaluated only on the test rows of that functional. Performance is then compared through the MAE on those stratified test subsets. This isolates the benefit of sharing information across XC labels from a change in the evaluation set.

\subsubsection*{Evaluation and metrics}

Model performance was evaluated on the material-grouped test sets of the five splits. Unless stated differently, we report the mean and sd of each metric over the five splits. The parity plots show the pooled test predictions of all splits, while their $R^{2}_{E}$ values are means over the splits. Each split metric describes a single split model trained on about 80\,\% of the materials and the sd describes the spread between splits. We report the mean absolute error (MAE) and the coefficient of determination ($R^{2}$) on the predicted correction~$\Delta$, together with the MAE of the raw DFT gaps with respect to experiment as a baseline. Because
$\hat{E}_{\mathrm{corr}}=E_{\mathrm{DFT}}+\hat{\Delta}$, the MAE on~$\Delta$ is equivalent to the MAE of the corrected gaps against experiment. 
%Importantly, here we report the $R^{2}_{E}$} for the corrected values, where we have a comparison between the $E_{\mathrm{corr}}$ and $E_{\mathrm{exp}}$. However, the $R^{2}_{\Delta}$} for the model is also reported, which is} computed with true and predicted values of $\Delta$.
The metrics were computed both in the train and test sets to compare training and test errors for every model built here (the corresponding table can be found in Supplementary Section S4). The variability from network initializations was assessed by evaluating each of the ten ensemble members separately (see Supplementary Section S8).

The mean absolute error is
\begin{equation}
\mathrm{MAE}
=
\frac{1}{N}\sum_{n=1}^{N}
\left|
\Delta_{n}-\hat{\Delta}_{n}
\right|
\end{equation}

Two coefficients of determination are used. The model score, $R^{2}_{\Delta}$, is evaluated on the correction,
\begin{equation}
R^{2}_{\Delta}
=
1-
\frac{
\sum_{n=1}^{N}
\left(\Delta_{n}-\hat{\Delta}_{n}\right)^{2}
}{
\sum_{n=1}^{N}
\left(\Delta_{n}-\bar{\Delta}\right)^{2}
},
\end{equation}
where $\bar{\Delta}=N^{-1}\sum_{n}\Delta_{n}$, and measures the fraction of the variance of the DFT error that the model explains.

In the parity plots, the agreement between calculated and experimental bandgaps is given by $R^{2}_{E}$, evaluated in the bandgap space,
\begin{equation}
R^{2}_{E}
=
1-
\frac{
\sum_{n=1}^{N}
\left(E_{\mathrm{exp},n}-E_{n}\right)^{2}}
{
\sum_{n=1}^{N}
\left(E_{\mathrm{exp},n}-\bar{E}_{\mathrm{exp}}\right)^{2}
},
\end{equation}
where $E_{n}$ is either the raw DFT gap ($E_{\mathrm{DFT},n}$), or the corrected gap ($E_{\mathrm{corr},n}$); and $\bar{E}_{\mathrm{exp}}=N^{-1}\sum_{n}E_{\mathrm{exp},n}$. The numerators of $R^{2}_{\Delta}$ and
$R^{2}_{E}$ coincide, but the denominators differ, so the two scores
are not the same.

Here, we also analyzed the relevance of each feature to the prediction through permutation importance \cite{breiman2001random} and Shapley Additive Explanations (SHAP) analysis \cite{lundberg2017unified}. For the NN and RF, we follow the standard procedure for permutation importance, i.e., the values of one descriptor are randomly exchanged between the test entries, the corrections are predicted again, and the increase of the test MAE is taken as the importance of that descriptor (a mean over 30 repetitions). The values were exchanged only among entries of the same XC functional, except for the XC label itself, which was exchanged over all test entries. The SHAP method, applied to the RF only, writes each predicted correction as a sum of contributions from the descriptors, based on the Shapley values of cooperative game theory~\cite{shapley1953value}. Family-level contributions are the sums over the descriptors of each family, averaged over the five splits. Details are given in Supplementary Section S3.

\subsubsection*{Model transferability}

To test the model transferability beyond the training distribution, the five split models were applied to infer the DFT error of the external QP-G$_0$W$_0$ benchmark of Großmann et al.~\cite{Grossmann2026}. The five splits serve to estimate the performance, and for new materials, the mean prediction of the five split models is used.

For each material, the required features were generated and missing descriptors (such as DOS or BS, when unavailable from MP) were filled by the median value from the training fit. Since some benchmark materials belong to the strict cohort, each material was labeled per split as \textit{in training} or \textit{not in training}, a material counting as in training only when both its composition and its crystal structure are in that split's training data. Fig.~\ref{fig:gw_comp} uses all benchmark materials in the bandgap window, with the mean of the five split models' predictions.  The transferability to materials outside the training set is evaluated per split on its materials \textit{not in training} (test or never-seen materials), with that split's model, and reported in the SI (see Supplementary Section S6). Notably, the NN was trained on a range for $\Delta$ of about $-1.7$ to $+2.4$\,eV, so it can be expected that the model may fail to accurately infer $\Delta$ for materials that lie outside that range, which are typically large-gap materials. Also, one should note that the descriptors considered here are strongly dependent on the atomic elements, and inference for elements that were not included in the training set might be another failure mode for the model (see Supplementary Section S6 for the elemental distribution of the training set). Therefore, we restricted the comparison with the QP-G$_0$W$_0$ benchmark to experimental bandgaps up to 8\,eV, as most semiconductors with potential applications lie within that range.

We use the MAE to compare the NN performance for corrected PBE and HSE06 gaps against experimental references and published QP-G$_0$W$_0$ gaps. Here, we report the metric for the bandgap window $[0,8]$\,eV, but the metric depends on the chosen range. A complete table with the dependence of the metrics on the bandgap range can be found in Supplementary Section S6.

Some of the materials in this benchmark were already included in the set used to train the model, while others were in the test set and others were unseen. Here, we compute the metrics without removing the previously seen training materials, but we report the metrics separately for the materials inside and outside the training set of each split in Supplementary Section S6. Among our set, there are 85 materials that were not used to train any of the five split models. 
%On the 85 benchmark materials not used to train any split model, this average is slightly more accurate than a single split model. 
Interestingly, if we average the predictions for them and then compute the MAE, we get 0.49 and 0.40\,eV for corrected PBE and HSE06, against $0.51 \pm 0.03$ and $0.42 \pm 0.03$\,eV, if we compute the MAE per split and then average them. Therefore, the per-split metrics reported here are slightly conservative for predictions on new materials.

To establish the relative computational costs for the cost--accuracy comparison, Ref.~\cite{das_2025} was used to set an estimation of order-of-magnitude factors relative to PBE for HSE06 and QP-G$_0$W$_0$ calculations. Note also that the MAE reported for QP-G$_0$W$_0$ is an average of the calculations with and without spin-orbit coupling. More details on the benchmark subset and the metrics behavior as a function of the bandgap range are given in Supplementary Section S6.

\subsubsection*{Use of generative AI}

Claude Code (Anthropic) was used to assist in writing and revision of Python scripts for the data curation, feature generation, model training and creation of figures. The model used was mainly Claude Opus 5.5. The same tool was used for language revision in the process of writing the main manuscript and Supplementary Information.  All code, results and text were reviewed and verified by the authors, who take full responsibility for the content.

\backmatter

\bmhead{Data and code availability}

The code with the full workflow (data curation, matching, feature generation, feature selection, training and evaluation), the matched experimental--theoretical dataset, the strict and permissive cohorts and their five splits, the trained models and the data behind every figure are available at \url{https://github.com/pcostacarvalho/correctingaps}. The experimental and theoretical bandgaps were taken from the sources cited in the Methods. A web application that applies the trained models to a single material, given its formula and space group, is available at \url{https://correctingaps.streamlit.app/}.

\bmhead{Acknowledgements}

Computational resources were provided by the National Academic Infrastructure for Supercomputing in Sweden (NAISS and LUMI), funded by the Swedish Research Council.
O.E. acknowledges financial support from the Swedish Research Council (VR), the Knut and Alice Wallenberg Foundation (KAW-Scholar program).
O.E. and E. K. D.-Cz. acknowledge support from the Wallenberg Initiative Materials Science for Sustainability (WISE) funded
by the Knut and Alice Wallenberg Foundation,  as well as support from STandUPP and eSSENCE.
P.C.C. acknowledges support from the Knut and Alice Wallenberg Foundation (KAW-Scholar program).
U.S. acknowledges funding support from the Swedish e-Science Research Centre (SeRC). S.I.S. acknowledges support from the Swedish Research Council (VR) (2023-05247). U.S, S.I.S., E. K. D.-Cz. and O.E acknowledge the support from Olle Engkvist Foundation.

\bmhead{Author contributions}

P.C.C. collected necessary data, performed technical work in training the neural network and wrote the first draft of the paper.

A.M. collected necessary data, performed technical work in training the neural network.

E. K. D.-Cz. and U.S. collected necessary data, discussed the approach and results, and coauthored the manuscript.

O.E. and S.I.S. conceived the project, discussed the approach and results, and coauthored the manuscript.

\bmhead{Competing interests}

The authors declare no competing interests.

%%===========================================================================================%%
%% For submission to npj journals via the eJP system, paste the reference list (from your    %%
%% .bbl file) directly into the manuscript .tex file instead of using \bibliography.          %%
%%===========================================================================================%%

\bibliography{sn-bibliography}% common bib file

\end{document}